\documentclass[prl,reprint,twocolumn,superscriptaddress,showpacs]{revtex4-1}

\usepackage{bm,mathrsfs}
\usepackage{graphicx}
\usepackage{epsfig}
\usepackage{amsmath,bbm}
\usepackage{amsfonts,amssymb}
\usepackage{times}
\usepackage{dsfont}
\usepackage{enumitem}  
\usepackage[colorlinks,linkcolor=blue,citecolor=blue]{hyperref}
\usepackage{comment}

\newcommand{\eqn}[1]{
\begin{eqnarray}
	#1
\end{eqnarray}
}

\begin{document}
\title{Solving quantum impurity problems in and out of equilibrium with variational approach}
\author{Yuto Ashida}
\email{ashida@cat.phys.s.u-tokyo.ac.jp}
\affiliation{Department of Physics, University of Tokyo, 7-3-1 Hongo, Bunkyo-ku, Tokyo 113-0033, Japan}
\author{Tao Shi} 
\email{tshi@itp.ac.cn}
\affiliation{CAS Key Laboratory of Theoretical Physics, Institute of Theoretical Physics, Chinese Academy of Sciences, P.O. Box 2735, Beijing 100190, China}
\affiliation{Max-Planck-Institut f\"ur Quantenoptik, Hans-Kopfermann-Strasse. 1, 85748 Garching, Germany}
\author{Mari Carmen Ba\~nuls}
\affiliation{Max-Planck-Institut f\"ur Quantenoptik, Hans-Kopfermann-Strasse. 1, 85748 Garching, Germany}
\author{J. Ignacio Cirac}
\affiliation{Max-Planck-Institut f\"ur Quantenoptik, Hans-Kopfermann-Strasse. 1, 85748 Garching, Germany}
\author{Eugene Demler}
\affiliation{Department of Physics, Harvard University, Cambridge, Massachusetts 02138, USA}

\date{\today}

\begin{abstract} 
A versatile and efficient variational approach is developed to solve in- and out-of-equilibrium problems of generic quantum spin-impurity systems. Employing the discrete symmetry hidden in spin-impurity models, we present a new canonical transformation that completely decouples the impurity and bath degrees of freedom. Combining it with Gaussian states, we present a family of many-body states to efficiently encode nontrivial impurity-bath correlations. We demonstrate its successful application to the anisotropic and two-lead Kondo models by studying their spatiotemporal dynamics and universal behavior in the correlations, relaxation times and the differential conductance. We compare them to previous analytical and numerical results. In particular, we apply our method to study new types of nonequilibrium phenomena that have not been studied by other methods, such as long-time crossover in the ferromagnetic easy-plane Kondo model. The present approach will be applicable to a variety of unsolved problems in solid-state and ultracold-atomic systems.
\end{abstract}


\maketitle

Understanding out-of-equilibrium dynamics of quantum many-body systems has become one of the central problems in physics. Recent experimental developments in diverse fields such as ultracold atoms \cite{EM11,CM12,FuT15,KAM16,RL17}, mesoscopic physics \cite{DFS02,THE11,LC11,IZ15,MMD17}, molecular electronics \cite{BDN11}, and carbon nanotubes \cite{MA07,CSJ12} have posed new theoretical questions for studying many-body dynamics driven by external fields or fast changes in the Hamiltonian. Quantum spin-impurity models (SIM), such as the famous Kondo model \cite{KJ64}, constitute a paradigmatic class of many-body systems which lie at the heart of many strongly correlated systems. Their nonequilibrium dynamics underly transport phenomena in mesoscopic systems \cite{GL88,NTK88,LW02,SF99,WWG00,RMP07,KAV11} and non-Fermi liquid behavior in heavy fermion materials \cite{HAC97,LH07,SQ10}, and give theoretical foundation for the real-time formulation of dynamical mean-field theory (DMFT)  \cite{GA96}. 

The ground-state properties of SIM are now well established by perturbative renormalization group (RG) \cite{APW70}, numerical renormalization group (NRG) \cite{WKG75} and the Bethe ansatz \cite{PBW80,AND80,ANL81,NK81,AN83,SP89}. The challenging and fascinating question of out-of-equilibrium  dynamics has recently come under active investigations in   theory \cite{AFB05,AFB06,AFB07,AFB08,RD08,JE10,LB14,WSR04,SP04,AHKA06,DSLG08,SH08,WA09,HMF09,HMF10,NHTM17,NM15,DB17,STL08,WP09,SM09,WP10,CG13,NP99,KA00,AR03,HA08,KM01,PM10,HA09L,HA09B,CT11,BS14,FS15,BCZ17,LF96,LF98,SA98,LD05,VR13,SG14,MM13,BCJ16} and experiments \cite{RL17,DFS02,THE11,LC11,IZ15,MMD17}. Examples include time-dependent NRG \cite{AFB05,AFB06,AFB07,AFB08,RD08,JE10,LB14}, density-matrix renormalization group (DMRG) \cite{WSR04,SP04,AHKA06,DSLG08,SH08,WA09,HMF09,HMF10,NHTM17}, time evolving block decimation (TEBD) \cite{NM15,DB17}, real-time Monte Carlo \cite{STL08,WP09,SM09,WP10,CG13}, perturbative RG \cite{NP99,KA00,AR03,HA08,KM01,PM10},  flow equation method \cite{HA09L,HA09B,CT11}, coherent-state expansion \cite{BS14,FS15,BCZ17}, and exact analyses  \cite{LF96,LF98,SA98,LD05,VR13,SG14,MM13,BCJ16}. Despite the rich variety of methods, they often become increasingly costly at long times due to, e.g., artifacts of the logarithmic discretization \cite{RA12} or large entanglement in the time-evolved state \cite{SU11}. Some of them can only determine the dynamics of the impurity but not that of the bath, or are restricted to particular parameter regimes. Moreover, it remains a major challenge to apply them to generic SIM beyond the simplest Kondo models. These challenges motivate the search for new approaches to quantum impurity systems.

\begin{figure}[b]
\includegraphics[width=86mm]{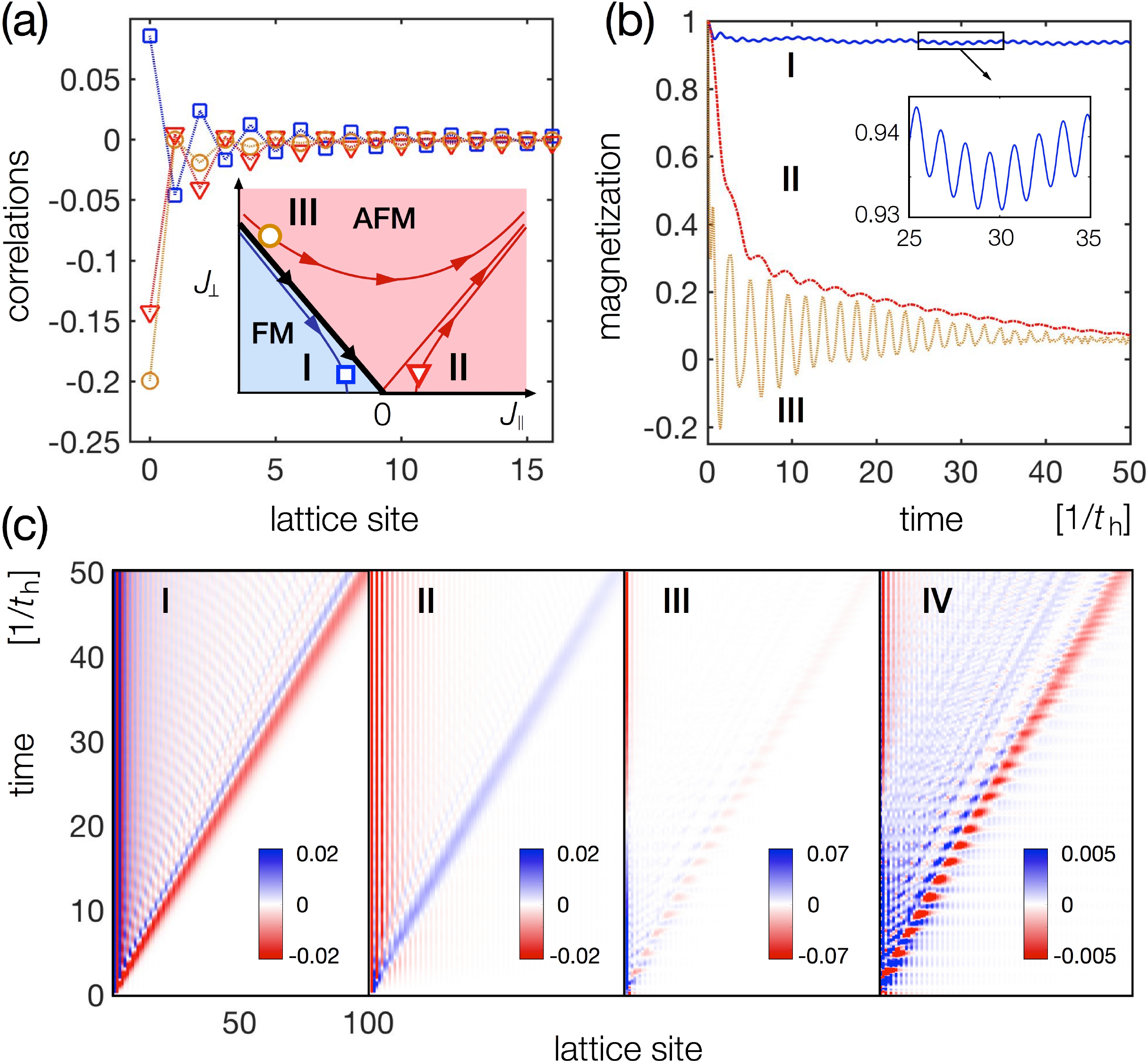} 
\caption{\label{fig_aniso}
(a) Ground-state impurity-bath spin correlation $\chi^{z}_{l}$ of the anisotropic Kondo model. (a,inset) The RG phase diagram and the parameters $(j_\parallel,j_\perp)$ corresponding to I~$(-0.5,0.2)$ (blue square) in the ferromagnetic phase (FM), II~$(0.5,0.2)$ (red triangle) and III~$(-1.85,2)$ (brown circle) in the antiferromagnetic phase (AFM). (b) Quench dynamics of the impurity magnetization $\langle\hat{\sigma}_{\rm imp}^{z}(t)\rangle$. (c) The corresponding spatiotemporal dynamics of correlations $\chi^{z}_{l}(t)$ in I~FM phase, II~AFM phase, III~easy-plane FM regime and IV~the same as in III but on a different scale. System size is $L=400$.
}
\end{figure}

In this Letter, introducing a new canonical transformation, we present a widely applicable variational approach to study in- and out-of-equilibrium properties of generic SIM. 
Besides the ability to efficiently capture the correct impurity-bath correlations and the conductance behavior, it reveals previously unexplored nonequilibrium dynamics such as ferromagnetic (FM) to antiferromagnetic (AFM) crossover (see the panels III and IV in Fig.~\ref{fig_aniso}c) in the FM easy-plane Kondo model. Such long-time spatiotemporal dynamics is difficult (if not impossible) to obtain in other approaches. Our versatile variational approach will pave the way towards solving interesting novel problems in both solid-state and ultracold-atomic systems.

{\it Canonical transformation.---} We first formulate our approach in the most general way as it is applicable to a wide class of SIM. The difficulty in SIM stems from the need to treat the strong entanglement between the impurity and bath. Here we introduce a new canonical transformation that completely decouples the impurity spin and bath degrees of freedom. We consider the Hamiltonian
\eqn{\label{totalH}
\hat{H}=\hat{H}_{\rm bath}+\hat{H}_{\rm int}+\hat{H}_{\rm imp},
}
where $\hat{H}_{\rm bath}=\sum_{lm\alpha}\hat{\Psi}^{\dagger}_{l\alpha}h_{lm}\hat{\Psi}_{m\alpha}$ describes an arbitrary single-particle Hamiltonian, with fermionic or bosonic creation (annihilation) operator $\hat{\Psi}^{\dagger}_{l\alpha}$ ($\hat{\Psi}_{l\alpha}$) for the $l$-th bath mode with spin $\alpha$. For simplicity, we consider a noninteracting spin-1/2 bath with $\alpha=\uparrow,\downarrow$ \footnote{A generalization of our variational approach to interacting bath having arbitrary bath spin is straightforward.}. The Hamiltonian $\hat{H}_{\rm int}=\hat{\bf s}_{\rm imp}\cdot\hat{\bf \Sigma}$ represents a generic interaction between the impurity and the bath
with $\hat{\bf s}_{\rm imp}=\hat{\boldsymbol \sigma}_{\rm imp}/2$ being the impurity spin-1/2 operator. We define the bath-spin operator including couplings as $\hat{\Sigma}^{\gamma}=\sum_{l}g_{l}^{\gamma}\hat{\sigma}_{l}^{\gamma}/2$ with $\hat{\sigma}_{l}^{\gamma}=\sum_{\alpha\beta}\hat{\Psi}_{l\alpha}^{\dagger}\sigma^\gamma_{\alpha\beta}\hat{\Psi}_{l\beta}$. The interaction strengths $g_{l}^{\gamma}$ are arbitrary and can be anisotropic and long-range. We also include the impurity Hamiltonian as $\hat{H}_{\rm imp}=-h_z\hat{s}_{\rm imp}^{z}$. Paradigmatic examples having the interaction form $\hat{H}_{\rm int}$ include the Kondo-type Hamiltonians \cite{KJ64} where the coupling $g_{l}^\gamma$ is local, and the central spin model \cite{JS03} where an interaction is long-range while $\hat{H}_{\rm bath}$ is frozen. 

To construct the canonical transformation, we observe that the Hamiltonian has a parity symmetry, $[\hat{H},\hat{\mathbb P}]=0$, with $\hat{\mathbb{P}}=\hat{\sigma}_{\rm imp}^{z}\hat{\mathbb P}_{\rm bath}$. Here, $\hat{\mathbb P}_{\rm bath}=e^{(i\pi/2)(\sum_{l}\hat{\sigma}^{z}_{l}+\hat{N})}$ is the parity operator acting on the bath, where $\hat{N}$ is the total particle number. The symmetry follows from the fact that $\hat{H}$ is invariant under the transformation $\hat{\mathbb P}^{-1}\,\hat{O}\,\hat{\mathbb P}$, which rotates the entire system around $z$ axis by $\pi$, i.e., transforms both impurity and bath spins as $\hat{\sigma}^{x,y}\to -\hat{\sigma}^{x,y}$. Our aim is to employ a parity conservation to find the disentangling transformation $\hat{U}$ satisfying $\hat{U}^{\dagger}\hat{\mathbb P}\hat{U}=\hat{\sigma}_{\rm imp}^{x}$ such that the impurity spin turns out to be a conserved quantity in the transformed frame. We can construct such a unitary transformation as
\eqn{\label{canonical}
\hat{U}=\exp\left[\frac{i\pi}{4}\hat{\sigma}_{\rm imp}^{y}\hat{\mathbb P}_{\rm bath}\right]=\frac{1}{\sqrt{2}}\left(1+i\hat{\sigma}_{\rm imp}^{y}\hat{\mathbb P}_{\rm bath}\right),
}
where we use $\hat{\mathbb P}_{\rm bath}^2=1$.
This leaves $\hat{H}_{\rm bath}$ invariant, while it maps the interaction onto $\hat{\tilde{H}}_{\rm int}=\hat{U}^{\dagger}\hat{H}_{\rm int}\hat{U}$:
\eqn{\label{transformed}
\hat{\tilde{H}}_{\rm int}\!=\!\hat{s}_{\rm imp}^x \hat{\Sigma}^{x}+\hat{\mathbb P}_{\rm bath}\left(-i\hat{\Sigma}^y/2+\hat{s}^x_{\rm imp}\hat{\Sigma}^z\right),
}
and $\hat{H}_{\rm imp}$ onto $\hat{\tilde{H}}_{\rm imp}=-h_z\hat{s}_{\rm imp}^{x}\hat{\mathbb P}_{\rm bath}$.
Remarkably, the impurity spin now commutes with the transformed Hamiltonian $[\hat{\tilde{H}},\hat{s}_{\rm imp}^x]=0$ and is thus completely decoupled from the bath degrees of freedom. The construction of $\hat{U}$ holds true for arbitrary conserved parity operator and can be readily applied to a variety of SIM, including two-impurity systems \cite{SP18}.

{\it Variational approach.---}
We combine the transformation~(\ref{canonical}) with fermionic Gaussian states \cite{WC12,CVK10} and introduce variational states to efficiently encode nonfactorizable impurity-bath correlations. A Gaussian state for the bath, $|\Psi_{\rm b}\rangle$, is completely determined by its covariance matrix $\Gamma$ \cite{WC12}:
\eqn{
(\Gamma)_{\xi l\alpha,\eta m\beta}=\frac{i}{2}\langle\Psi_{\rm b}|[\hat{\psi}_{\xi,l\alpha},\hat{\psi}_{\eta,m\beta}]|\Psi_{\rm b}\rangle,
}
where we introduce the Majorana operators $\hat{\psi}_{1,l\alpha}=\hat{\Psi}^{\dagger}_{l\alpha}+\hat{\Psi}_{l\alpha}$ and $\hat{\psi}_{2,l\alpha}=i(\hat{\Psi}^{\dagger}_{l\alpha}-\hat{\Psi}_{l\alpha})$. 
For the total system, we construct states of the form $|\Psi_{\rm tot}\rangle=\hat{U}|+_{x}\rangle_{\rm imp}|\Psi_{\rm b}\rangle$ with $\Gamma$ as variational parameters. Employing the time-dependent variational principle \cite{JR79,ST17}, we obtain the imaginary- and real-time evolution equations for $\Gamma$:
\eqn{\label{imag}
\frac{d\Gamma}{d\tau}&=&-{\cal H}-\Gamma {\cal H}\Gamma,\\\label{real}
\frac{d\Gamma}{dt}&=&{\cal H}\Gamma-\Gamma {\cal H},
}
where ${\cal H}=4\delta E/\delta\Gamma$ is the functional derivative  of the mean energy  $E=\langle\Psi_{\rm tot}|\hat{H}|\Psi_{\rm tot}\rangle$  \cite{YA18F}. The variational ground state can be obtained in the limit $\tau\to\infty$ in the imaginary-time evolution~(\ref{imag}). In contrast, Eq.~(\ref{real}) allows us to calculate the real-time dynamics of SIM. 

{\it Equilibrium properties.---}
As a paradigmatic example, we first apply our approach to the anisotropic Kondo model:
\eqn{
\hat{H}&=&-t_{\rm h}\sum_{l=-L}^{L}\left(\hat{c}^{\dagger}_{l,\alpha}\hat{c}_{l+1,\alpha}+{\rm h.c.}\right)\nonumber\\
&&\!+\!\frac{J_{\perp}}{4}\!\!\sum_{\gamma=x,y}\!\!\hat{\sigma}^{\gamma}_{\rm imp}\hat{c}_{0,\alpha}^{\dagger}\sigma^{\gamma}_{\alpha\beta}\hat{c}_{0,\beta}\!+\!\frac{J_\parallel}{4}\hat{\sigma}^{z}_{\rm imp}\hat{c}_{0,\alpha}^{\dagger}\sigma^{z}_{\alpha\beta}\hat{c}_{0,\beta},
}
where $\hat{c}_{l,\alpha}^{\dagger}$ ($\hat{c}_{l,\alpha}$) creates (annihilates) a fermion with position $l$ and spin $\alpha$, the summations over $\alpha,\beta$ are contracted. We denote the dimensionless Kondo couplings as $j_{\parallel,\perp}=\rho_{\rm F}J_{\parallel,\perp}$ with $\rho_{\rm F}=1/(2\pi t_{\rm h})$ being the density of states at the Fermi energy. We choose the unit $t_{\rm h}=1$ hereafter.

The anisotropic Kondo model exhibits a quantum phase transition \cite{LAJ87} between FM and AFM phases as shown in the RG phase diagram \cite{APW70} in the inset of Fig.~\ref{fig_aniso}a. In the main panel, we show the ground-state impurity-bath spin correlations $\chi^{z}_{l}=\langle\hat{\sigma}^{z}_{\rm imp}\hat{\sigma}_{l}^{z}\rangle/4$ in three different regimes.  The FM results at I (blue square) and AFM results at II (red triangle) indicate the formation of the triplet and singlet pairs of the impurity and bath spins, respectively. Importantly, our method also correctly reveals the AFM nature at III (brown circle) that is close to the phase boundary. 

\begin{figure}[t]
\includegraphics[width=86mm]{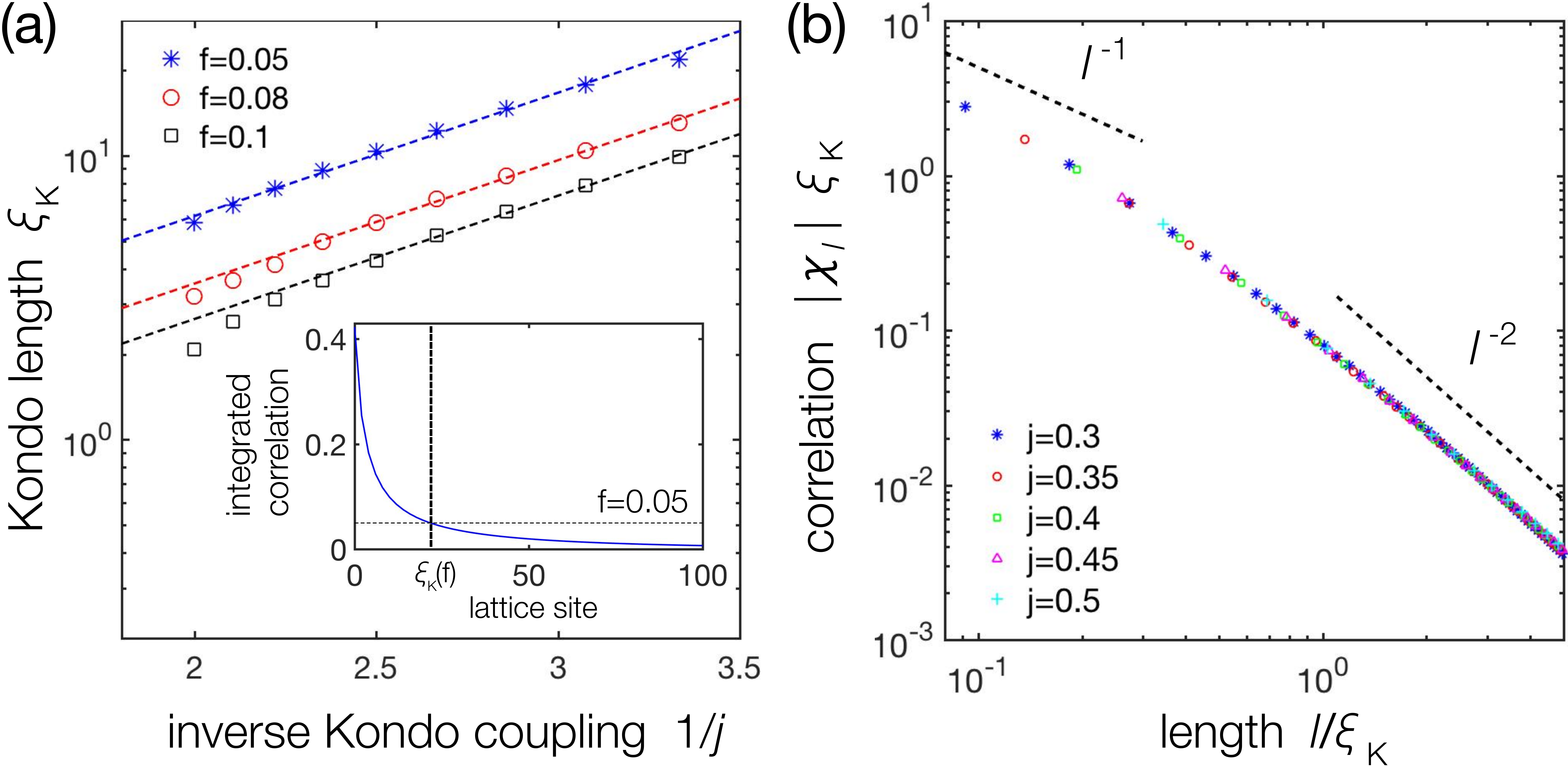} 
\caption{\label{fig_gs}
Ground-state properties of the Kondo model. (a) Screening length $\xi_{\rm K}$ plotted for different Kondo coupling $j=\rho_{\rm F}J$ and thresholds $f$. The dashed lines indicate the scaling $\xi_{\rm K}\propto e^{1/j}$. (inset) The Kondo length $\xi_{\rm K}$ is extracted as a length scale in which a fraction $1-f$ of total antiferromagnetic correlations is contained. (b) Spin correlations plotted in the dimensionless unit of $\xi_{\rm K}$ for $f=0.05$, collapsing onto the universal curve. The dashed lines indicate the scaling $l^{-1}$ ($l^{-2}$) in short (long) distance. System size is $L=400$.}
\end{figure}

As a critical test of our approach, we extract the Kondo screening length $\xi_{\rm K}$ in the variational ground state and test the universal behavior in the SU(2)-symmetric case $j=j_\parallel=j_\perp>0$. 
We determine $\xi_{\rm K}$ as the length scale below which most of the Kondo screening cloud is developed \cite{AH09,NM15}. Specifically, 
we introduce a threshold $f$ for the integrated antiferromagnetic
correlations $\Sigma_{\rm AF}(l)=\sum_{|m|=0,2,4\ldots}^{l}\chi_{m}$ (Fig.~\ref{fig_gs}a, inset) with $\chi_{m}=\langle\hat{\boldsymbol \sigma}_{\rm imp}\cdot\hat{\boldsymbol \sigma}_{m}\rangle/4$, and extract $\xi_{\rm K}$ from the implicit relation: $f=1-\Sigma_{\rm AF}(\xi_{\rm K}(f))/\Sigma_{\rm AF}(L)$ \footnote{Here we sum the correlations over even sites only to avoid cancellations from ferromagnetic contributions on odd sites and obtain a better accuracy of fitting procedure to extract Kondo length $\xi_{\rm K}$}. Figure~\ref{fig_gs}a plots the extracted $\xi_{\rm K}(f)$ against the inverse Kondo coupling $1/j$ for different $f$. The results agree with the nonperturbative scaling  $\xi_{\rm K}\propto T_{\rm K}^{-1}\propto e^{1/j}$ \cite{HAC97} independent of the choice of $f$. As a further test, we plot $\chi_l$ in units of the extracted $\xi_{\rm K}$ (Fig.~\ref{fig_gs}b). Remarkably, all the results for different Kondo couplings $j$ collapse onto the same universal curve and show the crossover from $l^{-1}$ to $l^{-2}$ decay at $l/\xi_{\rm K}\sim 1$ \cite{IH78,BV98,HT06}. 
To avoid finite-size and lattice effects, here we set $j$ large enough such that $\xi_{\rm K}\ll L$  while it is kept small enough so that $\xi_{\rm K}$ is still larger than the lattice constant. To meet the former condition, we ensure that the sum rule $\sum_l\chi_{l}=-3/4$  \cite{BL07} is satisfied with an error below $0.5$\%. 

{\it Out-of-equilibrium dynamics.---}
We now apply our approach to study out-of-equilibrium dynamics. To be concrete, we analyze the quench dynamics starting from the initial state $|\!\!\uparrow\rangle_{\rm imp}|{\rm FS}\rangle$, where $|{\rm FS}\rangle$ represents the half-filled Fermi sea of the bath. Previously, using the bosonization mapping between the Kondo model and the spin-boson model \cite{LAJ87}, the relaxation dynamics have been studied by NRG \cite{AFB06} and the bosonic Gaussian states combined with a unitary transformation \cite{ST17}. While the latter has been specifically designed to the spin-boson model, our approach is applicable to generic SIM. Moreover, in the previous methods one had to use strictly linear dispersion and to introduce an artificial cut-off energy. Hence, one of distinctive features in our approach is that it can be applied without relying on the bosonization and thus allows for a quantitative comparison with an experimental system. This is particularly important in light of recent experimental developments in simulating dynamics of SIM \cite{RL17,DFS02,THE11,LC11,IZ15,MMD17,BDN11,KM17}. 

Figures~\ref{fig_aniso}b,c show the magnetization dynamics $\langle\hat{\sigma}_{\rm imp}^{z}(t)\rangle$ and spatiotemporal spreading of spin correlations $\chi^{z}_{l}(t)$ after the quench. As shown in the panels I and II in Fig.~\ref{fig_aniso}c, spin correlations develop FM and AFM correlations after passing through the ``light cone" created by AFM and FM ballistic spin waves, respectively. These AFM (FM) spin waves result from the excess spin in the generation of the triplet (singlet) pair around the impurity. As shown in Fig.~\ref{fig_aniso}b, the magnetization eventually relaxes to a value close to zero in the AFM phase, indicating the formation of the Kondo singlet, while the value remains finite in the FM phase. The dynamics associate with the fast oscillations having period characterized by the bandwidth $2\pi/{\cal D}$ with ${\cal D}=4t_{\rm h}$ and $\hbar=1$ (see e.g., Fig.~\ref{fig_aniso}b inset). These fast oscillations originate from  high-energy excitations of a particle from the bottom of the band \cite{KM12} and were absent in the bosonized treatments. 

Most interestingly, at the point III in easy-plane FM regime ($|J_\parallel|<|J_\perp|$), spin correlations exhibit the distinct crossover dynamics from FM to AFM (panel III in Fig.~\ref{fig_aniso}c). As shown in the closeup panel IV, the initial development of FM correlations leads to the emission of ballistic AFM spin waves while the subsequent crossover to AFM associates with the repeated emissions of FM spin waves. The origin of such crossover can be understood from the nonmonotonic RG flows in this regime (Fig.~\ref{fig_aniso}a, inset), where short (long) time dynamics is governed by the high (low) energy physics characterized by FM (AFM) coupling $J_\parallel$ ($J_\perp$). Here, the real time effectively plays the role of the inverse RG scale \cite{NP99}. The predicted spatiotemporal dynamics can be readily tested with site-resolved measurements as allowed by quantum gas microscopy \cite{EM11,CM12,FuT15,KAM16}.

\begin{figure}[b]
\includegraphics[width=86mm]{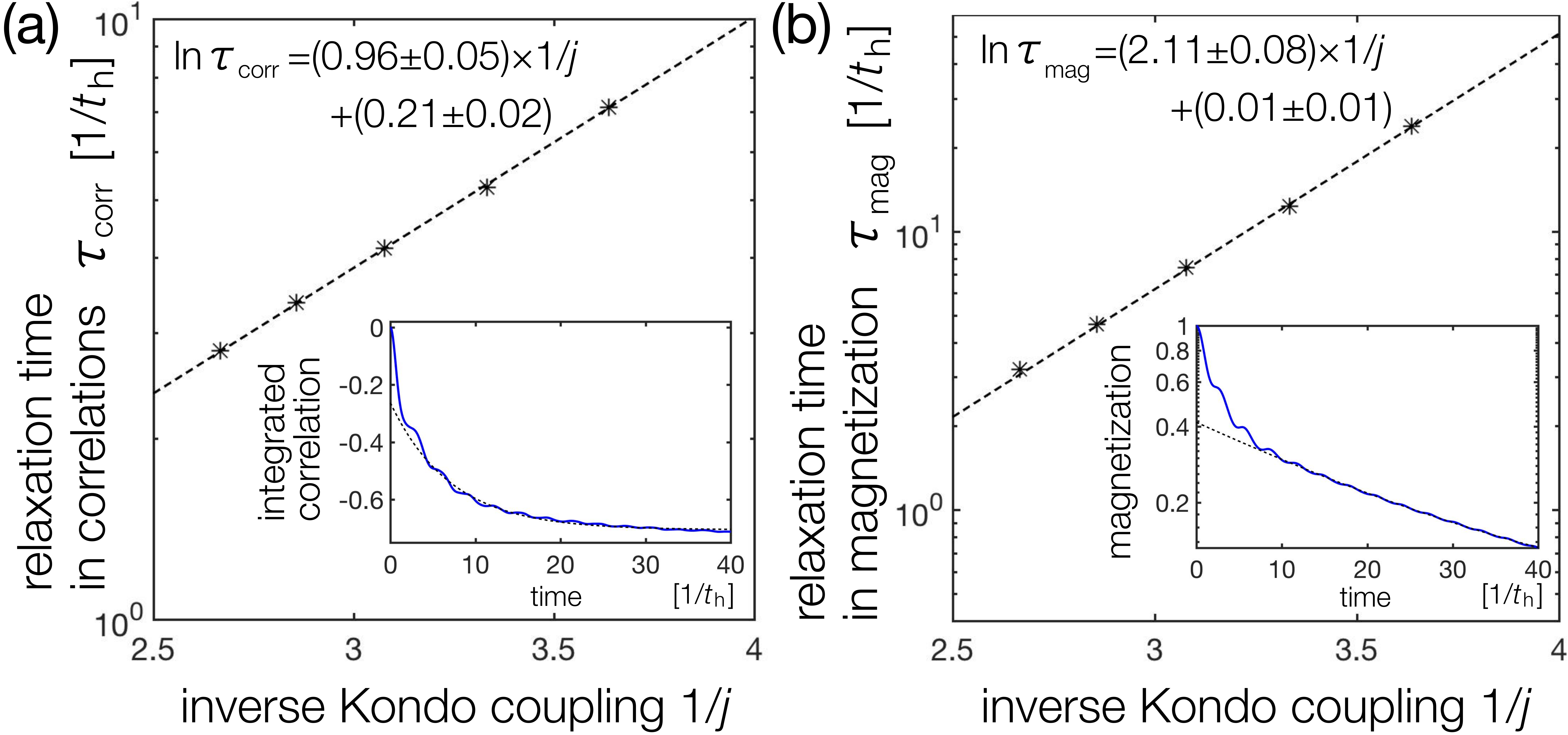} 
\caption{\label{fig_dy}
Relaxation time scales $\tau$ in (a) correlation and (b) magnetization plotted for different Kondo coupling $j=\rho_{\rm F}J$. (Insets) The relaxation times are extracted by fitting  $\Sigma_{\rm AF}(L)(t)$ and $\langle\hat{\sigma}_{\rm imp}^{z}(t)\rangle$ in long-time regime with the function $a+b e^{-t/c}$. The dashed lines in the main panels indicate the fitted lines, showing nonperturbative scaling $\ln\tau\propto 1/j$. System size is $L=400$.}
\end{figure}

As a critical test, we study the nonperturbative scalings of the relaxation time scales $\tau_{\rm corr}$ for the integrated correlations $\Sigma_{\rm AF}(L,t)$ and $\tau_{\rm mag}$ for the impurity magnetization $\langle\hat{\sigma}^{z}_{\rm imp}(t)\rangle$ in the SU(2)-symmetric case.
After the quench, each observable eventually relaxes to its steady-state value and we extract the relaxation times by fitting the tale dynamics with an exponential function (Fig.~\ref{fig_dy}a,b inset). The main panels show that within numerical errors the relaxation times for both observables show the nonperturbative dependence  $\tau_{\rm corr}\propto e^{1/j}$ and $\tau_{\rm mag}\propto e^{2/j}$. The observed different scalings agree with the TEBD results \cite{NM15}: $ \tau_{\rm corr}\propto e^{(1.5\pm 0.2)/j}$ and $\tau_{\rm mag}\propto e^{(1.9\pm 0.2)/j}$ (a rather large deviation in $\tau_{\rm corr}$ has been attributed to the difficulty of taking the adiabatic limit in the Anderson model).


{\it Transport dynamics.---}
\begin{figure}[t]
\includegraphics[width=86mm]{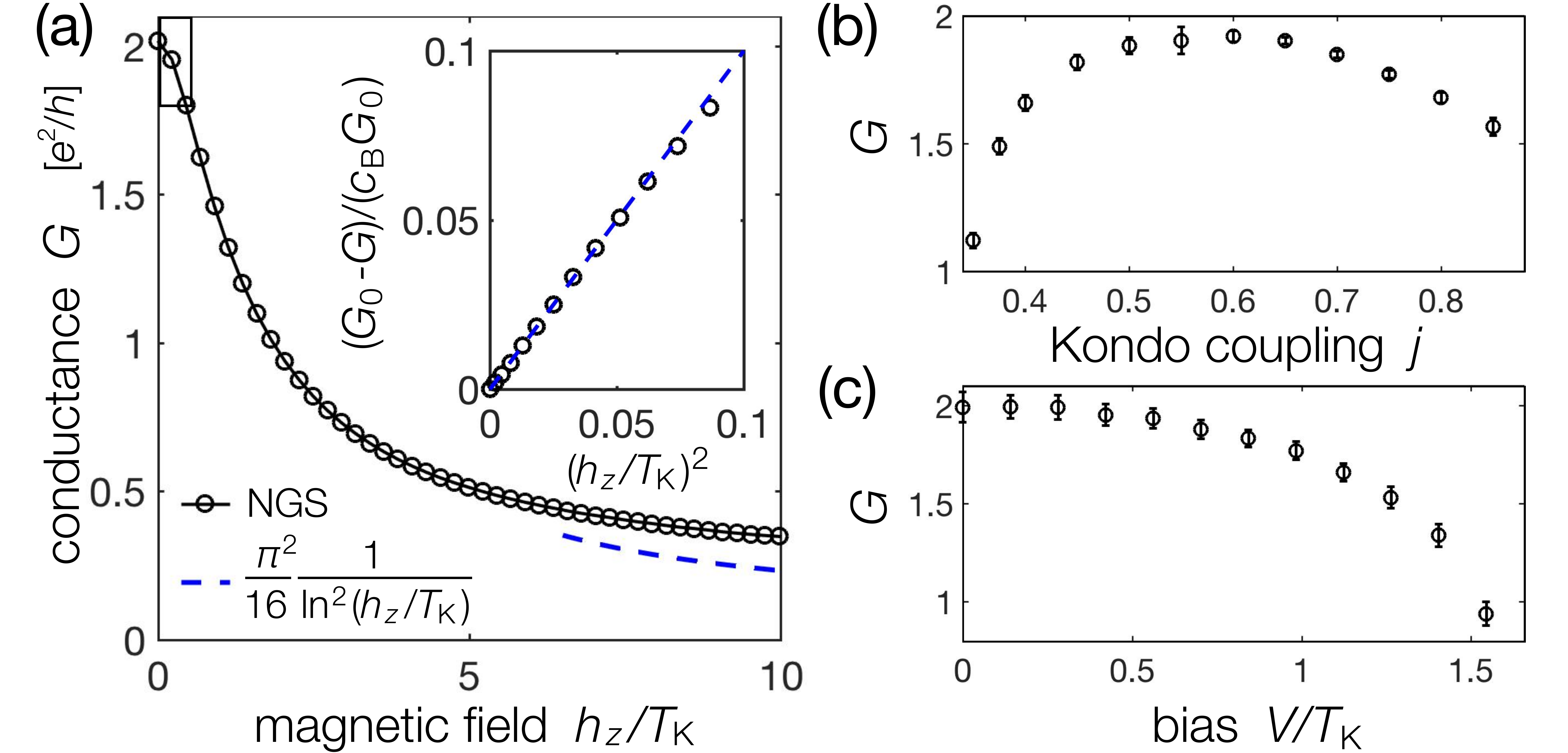} 
\caption{\label{fig_trans}
Differential conductance $G$ with varying (a) magnetic field $h_z/T_{\rm K}$, (b) Kondo coupling $j$, and (c) bias potential $V/T_{\rm K}$. In (a), we show the obtained results (black open circles) and the asymptotic scalings with the infinite-bandwidth approximation (blue dashed lines). The inset magnifies the low-field behavior. Kondo temperature $T_{\rm K}$ is extracted from the magnetic susceptibility $\chi=1/(4T_{\rm K})$. System size is $L=200$ for each lead and we use (a) $j=0.35$ and $V=0$, (b) $h_z=0$ and $V=0.8t_{\rm h}$, and (c) $h_z=0$ and $j=0.4$.}
\end{figure}
We finally apply our approach to  the two-lead Kondo model \cite{KA00} that is relevant to experiments in mesoscopic systems. We consider the Hamiltonian
\eqn{\label{two-leads}
\hat{H}_{\rm two}&=&\sum_{l\eta}\biggl[-t_{\rm h}\bigl(\hat{c}^{\dagger}_{l,\alpha\eta}\hat{c}_{l+1,\alpha\eta}\!+\!{\rm h.c.}\bigr)+eV_{\eta}\,\hat{c}^{\dagger}_{l,\alpha\eta}\hat{c}_{l,\alpha\eta}\biggr]\nonumber\\
&&+\frac{J}{4}\sum_{\eta\eta'}\hat{\boldsymbol{\sigma}}_{\rm imp}\cdot\hat{c}_{0,\alpha\eta}^{\dagger}\boldsymbol{\sigma}_{\alpha\beta}\hat{c}_{0,\beta\eta'},
}
where $\eta={\rm L,R}$ denotes the left (L) or right (R) lead. We set the bias potential $V_{\rm L,R}$ of each lead to be $V_{\rm L,R}=\pm V/2$. The initial condition is $|\!\!\uparrow\rangle_{\rm imp}|{\rm FS}\rangle_{\rm L}|{\rm FS}\rangle_{\rm R}$ with $|{\rm FS}\rangle_{\rm L,R}$ being the half-filled Fermi sea of each lead. We then quench the Hamiltonian (\ref{two-leads}) and study the dynamics of the current $I(t)$ between the two leads:
\eqn{
I(t)=\frac{ie}{4\hbar}J\sum_{\alpha\beta}\left[\langle\hat{\boldsymbol \sigma}_{\rm imp}\cdot\hat{c}_{0\alpha {\rm L}}^{\dagger}\boldsymbol{\sigma}_{\alpha\beta}\,\hat{c}_{0\beta {\rm R}}\rangle-{\rm h.c.}\right].
} 
After the quench, the current eventually reaches its quasi-steady value. We determine the differential conductance $G=d\overline{I}/dV$ from the steady current $\overline{I}(V)$obtained by taking time average \cite{AHKA06,DSLG08,HMF09,JE10}. Applying a magnetic field $h_z$, we confirm the quadratic behavior $G_{0}(1-c_{B}(h_z/T_{\rm K})^2)$ with the correct coefficient $c_{B}=\pi^2/16$ at low field and the logarithmic behavior  $\pi^2G_{0}/(16\ln^2(h_z/T_{\rm K}))$ at high field, where $G_0$ is the conductance at the zero field \cite{AR01,ACH05,GL05,SE09,MC09,KAV11,MC15,FM17,OA18,OA182} (Fig.~\ref{fig_trans}a).
In contrast, if we change the Kondo coupling $j$, we expect the nonmonotonic behavior because $G$ is trivially zero at $j=0$, while it should degrade in $j\to\infty$ due to the formation of the bound state tightly localized at the impurity site, which prevents other electrons from approaching the junction. Figure~\ref{fig_trans}b confirms this nonmonotonic dependence of $G$ against the Kondo coupling $j$. Different from two-channel systems \cite{MAK12,MAK16,MAK17}, the nonmonotonicity originates from intrinsically finite bandwidth in the lattice model and is absent in the conventional infinite-bandwidth treatment \cite{AR01,SN162}. 

 Figure~\ref{fig_trans}c shows the nonlinear conductance behavior at finite bias $V$. Two remarks are in order. Firstly, the numerical error due to current fluctuation in time obscures minuscule changes of $G$ in $V\ll T_{\rm K}$, making it difficult to precisely test the quadratic behavior \cite{AHKA06,DSLG08,HMF09,JE10} in the perturbative regime. This can be worked out if we implement our approach in a different way based on the linear response theory.  Secondly, in $V\gg T_{\rm K}$ the bias eventually becomes comparable to the finite bandwidth (and to the Fermi energy) and calculations of current and conductance are no longer faithful. This is a common limitation in real-space calculations \cite{HMF09,JE10} and can be avoided if one uses the momentum basis of bath modes and specify the linear dispersion with a large bandwidth. Yet, we emphasize that the present implementation is already reliable (at least) in the intermediate regime $V\sim T_{\rm K}$.


{\it Discussions.---}
A simple entanglement-based argument can give insights into the success of our approach. On the one hand, our variational approach considers the following family of states:
\eqn{
|\Psi_{\rm tot}\rangle&=&\hat{U}|+_{x}\rangle_{\rm imp}|\Psi_{\rm b}\rangle\nonumber\\
&=&|\!\uparrow~\!\!\rangle_{\rm imp}\hat{\mathbb P}_{+}|\Psi_{\rm b}\rangle+|\!\downarrow~\!\!\rangle_{\rm imp}\hat{\mathbb P}_{-}|\Psi_{\rm b}\rangle,
\label{tot_entanglement}
}
where $|\Psi_{\rm b}\rangle$ is a Gaussian state and $\hat{{\mathbb P}}_{\pm}=(1\pm\hat{{\mathbb P}}_{\rm bath})/2$. On the other hand, a recent study \cite{YC17} has shown that most of the entanglement in the Kondo singlet takes place with just one specific single-particle state, leading to the approximative expression originally suggested by Yosida \cite{KY66}:
\eqn{\label{kondo_entanglement}
|\Psi_{\rm Kondo}\rangle\!=\!\frac{1}{\sqrt 2}\!\left(|\!\uparrow~\!\!\rangle_{\rm imp}\,\hat{d}^{\dagger}_{\downarrow}|{\rm FS}\rangle\!-\!|\!\downarrow~\!\!\rangle_{\rm imp}\,\hat{d}^{\dagger}_{\uparrow}|{\rm FS}\rangle\right),
}
where $\hat{d}^{\dagger}_{\sigma}=\sum_{l}d_{l}\hat{c}_{l\sigma}^{\dagger}$ is the dominant single-particle state.
In fact, Eq.~(\ref{kondo_entanglement}) belongs to our family of variational states (\ref{tot_entanglement}) as shown by the choice $|\Psi_{\rm b}\rangle=(\hat{d}^{\dagger}_{\downarrow}-\hat{d}^{\dagger}_{\uparrow})|{\rm FS}\rangle/\sqrt{2}$. This observation indicates the ability of our variational state to efficiently encode the most significant part of the impurity-bath entanglement. Yet, we stress that our variational states go beyond the simple ansatz~(\ref{kondo_entanglement}) since they take into account general Gaussian states instead of the trivial Fermi sea. Such a flexibility is crucial to obtain quantitatively accurate results \cite{YA18F}.

In summary, we presented a versatile and efficient variational approach to study in- and out-of-equilibrium physics of SIM. Despite its simplicity, we demonstrated in the anisotropic and two-lead Kondo models that the variational states successfully capture the correct correlations and conductance behavior. In particular, it has already found applications to revealing previously unexplored physics such as the long-time crossover dynamics. Further details can be found in the accompanying paper \cite{YA18F}, where the full expression of the functional derivative $\cal H$ and the benchmark results with the matrix-product-state calculations are presented. 

The present approach should be applicable to a variety of interesting unsolved problems in both solid-state and ultracold-atomic systems. 
For instance, our approach can be readily generalized to bosonic systems \cite{FGM04,FS06,FFM11,FT15}, the Anderson model and multiple impurities \cite{SP18}, which will be published elsewhere. Another promising direction is an extension of our approach to multi-channel systems \cite{RMP07,IZ15,MAK12,MAK16,MAK17} and the central spin model \cite{JS03}.
A generalization to finite temperatures is possible by using Gaussian density matrices. Including the phase factor, it is also possible to calculate the spectral function \cite{AR03,WA09}. 
It is particularly interesting to test the maximally fast information scrambling \cite{MJ16} in the non-Fermi liquid phase of the multi-channel Kondo models \cite{DB17}. On another front, the proposed variational approach could be applied as a basis for a new type of impurity solver for DMFT \cite{GA96}. 

\paragraph*{Acknowledgements.---} We acknowledge Carlos Bolech Gret, Adrian E. Feiguin, Shunsuke Furukawa, Leonid Glazman, Vladimir Gritsev, Masaya Nakagawa and Achim Rosch for fruitful discussions. Y.A. acknowledges support from the Japan Society for the Promotion of Science through Program for Leading Graduate Schools (ALPS) and Grant No.~JP16J03613, and Harvard University for hospitality. T.S. acknowledges the Thousand-Youth-Talent Program of China. J.I.C. is supported by the ERC QENOCOBA under the EU Horizon 2020 program (grant agreement 742102).
E.D. acknowledges support from Harvard-MIT CUA, NSF Grant No. DMR-1308435, AFOSR Quantum Simulation MURI, AFOSR grant number FA9550-16-1-0323, the Humboldt Foundation, and the Max Planck Institute for Quantum Optics.

\bibliography{reference}

\end{document}